\documentclass[12pt]{iopart}
\usepackage{graphicx}
\usepackage{bm}
\usepackage{subfigure}
\begin{document}
\title{Improved cosmological constraints on the curvature and equation of state of dark energy}
\author{Nana Pan}
\address{College of Mathematics and Physics,
Chongqing University of Posts and Telecommunications, Chongqing 400065, China}
\ead{pannn@cqupt.edu.cn}
\author{Yungui Gong }
\address{College of Mathematics and Physics,
Chongqing University of Posts and Telecommunications, Chongqing 400065, China }
\ead{gongyg@cqupt.edu.cn }
\author{Yun Chen}
\address{Department of Astronomy, Beijing Normal University,
Beijing 100875, China}
\author{Zong-Hong Zhu}
\address{Department of Astronomy, Beijing Normal University,
Beijing 100875, China}
\ead{zhuzh@bnu.edu.cn}

\begin{abstract}
We apply the Constitution compilation of 397 supernova Ia, the
baryon acoustic oscillation measurements including the $A$
parameter, the distance ratio and the radial data, the five-year
Wilkinson microwave anisotropy probe and the Hubble parameter data
to study the geometry of the universe and the property of dark
energy by using the popular Chevallier-Polarski-Linder and
Jassal-Bagla-Padmanabhan parameterizations. We compare the simple
$\chi^2$ method of joined contour estimation and the Monte Carlo
Markov chain method, and find that it is necessary to make the
marginalized analysis on the error estimation. The probabilities
of $\Omega_k$ and $w_a$ in the Chevallier-Polarski-Linder model
are skew distributions, and the marginalized $1\sigma$ errors are
$\Omega_m=0.279^{+0.015}_{-0.008}$,
$\Omega_k=0.005^{+0.006}_{-0.011}$, $w_0=-1.05^{+0.23}_{-0.06}$,
and $w_a=0.5^{+0.3}_{-1.5}$. For the Jassal-Bagla-Padmanabhan
model, the marginalized $1\sigma$ errors are
$\Omega_m=0.281^{+0.015}_{-0.01}$,
$\Omega_k=0.000^{+0.007}_{-0.006}$, $w_0=-0.96^{+0.25}_{-0.18}$,
and $w_a=-0.6^{+1.9}_{-1.6}$. The equation of state parameter
$w(z)$ of dark energy is negative in the redshift range
$0\le z\le 2$ at more than $3\sigma$ level. The flat
$\Lambda$CDM model is consistent with the current observational
data at the $1\sigma$ level.
\end{abstract}
 \maketitle

\section{Introduction}
The accelerating expansion of the universe was first discovered by
the type Ia supernova (SN Ia) observations \cite{acc1,acc2}. The
phenomena of acceleration could be explained straightforwardly by
introducing an exotic source of matter with negative pressure, the
so-called dark energy, which dominates the total matter content of
the universe at the present epoch and causes the expansion to
accelerate. During the past decade, in addition to the simple
cosmological constant model, a lot of dynamical dark energy
models, such as the quintessence \cite{quint}, phantom
\cite{phantom}, k-essence \cite{k}, tachyon \cite{tachyonic},
quintom \cite{Feng:2004ad}, h-essence \cite{hessence}, Chaplygin
gas \cite{chaplygin}, holographic dark energy \cite{holo}, $f(R)$
\cite{fr}, Dvali-Gabadadze-Porrati \cite{dgp} models, etc, have
been proposed. Although a lot of efforts have been made to
understand the driving force of the accelerating expansion and the
property of dark energy, whether dark energy is dynamical or not
is still an open question. Therefore, it is necessary to study the
nature of dark energy such as the evolutions of its energy density
and equation of state.

Apart from phenomenological models, another effective approach to
study dark energy is through the observational data. Recently,
based on the popular Chevallier-Polarski-Linder (CPL)
parametrization of dark energy  \cite{cpl}, it was found that the
flat $\Lambda$CDM model is inconsistent with the current data at
more than $1\sigma$ level \cite{star,huang}. In \cite{star}, it
was suggested that the cosmic acceleration is slowing down from
$z\sim 0.3$. In \cite{huang}, it was claimed that dark energy
suddenly emerged at redshift $z\sim 0.3$. Furthermore, possible
oscillating behavior of dark energy was found in \cite{cai}.
However, no evidence for dark energy dynamics was found  in
\cite{corray9,gong,gong10}. It was argued that the systematics in
different data sets heavily affected the fitting results from
observational data \cite{gong,gong10}. To further study the
dynamics of dark energy, it is necessary to apply more
complimentary observational data. In this paper, we combine the
Constitution sample of 397 SN Ia data \cite{consta}, the model
independent $A$ parameter from the baryon acoustic oscillation
(BAO) measurements \cite{bar}, the two BAO distance ratios at
$z=0.2$ and $z=0.35$ \cite{wjp}, the radial BAO measurements at
$z=0.24$ and $z=0.43$ \cite{eg}, the five-year Wilkinson microwave
anisotropy probe data (WMAP5) \cite{kdn}, and the Hubble parameter
$H(z)$ data \cite{hz1,hz2} to probe the geometry of the universe
and the nature of dark energy by using the CPL and
Jassal-Bagla-Padmanabhan (JBP) \cite{jbp} parameterizations. We
first use the simple $\chi^2$ method of joined contour estimation
to obtain the constraints on the model parameters. However, the
simple $\chi^2$ method by fixing other parameters at their best
fit values has some drawbacks because we neglect the correlation
effects between the parameters and the degeneracy between
parameters was not considered. When the parameters are strongly
correlated, the errors of some parameters will be under-estimated
if we fix the other parameters at their best fit values. So we
also apply the Monte Carlo Markov chain (MCMC) method to obtain
the marginalized errors of the model parameters. The advantage of
the MCMC method is that it considers the correlations between the
model parameters and the result is more reliable.

The paper is organized as follows. In section 2, we present the SN
Ia data \cite{consta}, the BAO data \cite{bar,wjp,eg}, the WMAP5
data \cite{kdn} and the $H(z)$ data, and all the formulas related
with these data. In section 3, We use the $\Lambda$CDM model as an
example to show how to apply the data to constrain cosmological
models. In section 4, we use the CPL model to study the geometry
of the universe and the property of dark energy. The JBP model is
used to probe the geometry of the universe and the evolution of
dark energy in section 5. We conclude the paper in section 6.

\section{Fitting procedure}

To use the Constitution compilation of 397 SN Ia data \cite{consta},
we minimize
\begin{equation}
\label{chi}
\chi^2=\sum_{i=1}^{397}\frac{[\mu(z_i)-\mu_{obs}(z_i)]^2}{\sigma^2_i},
\end{equation}
where the extinction-corrected distance modulus
$\mu(z)=5\log_{10}[d_L(z)/{\rm Mpc}]+25$, $\sigma_i$ is the total
uncertainty in the SN Ia observation, the luminosity distance $d_L(z)$
is
\begin{equation}
\label{lum} d_L(z)=\frac{1+z}{H_0\sqrt{|\Omega_{k}|}} {\rm
sinn}\left[\sqrt{|\Omega_{k}|}\int_0^z \frac{dz}{E(z)}\right],
\end{equation}
the dimensionless Hubble parameter $E(z)=H(z)/H_0$, and
\begin{eqnarray}
\frac{{\rm sinn}(\sqrt{|\Omega_k|}x)}{\sqrt{|\Omega_k|}}=\left\{\begin{array}{lr}
\sin(\sqrt{|\Omega_k|}x)/\sqrt{|\Omega_k|},& {\rm if}\ \Omega_k<0,\\
x, & {\rm if}\  \Omega_k=0, \\
\sinh(\sqrt{|\Omega_k|}x)/\sqrt{|\Omega_k|}, & {\rm if}\  \Omega_k>0.
\end{array}\right.
\end{eqnarray}
Due to the arbitrary normalization of the luminosity distance, the nuisance parameter
$h$ in the SN Ia data is not the observed Hubble constant. So we marginalize the nuisance parameter $h$ with a flat prior,
after the marginalization, we get \cite{gong08},
\begin{equation}
\label{chi1} \chi^2_{sn}(\mathbf{p})=\sum_{i=1}\frac{\alpha_i^2}{\sigma^2_i}-\frac{(\sum_i\alpha_i/\sigma_i^2-\ln 10/5)^2}{\sum_i 1/\sigma_i^2}
-2\ln\left(\frac{\ln 10}{5}\sqrt{\frac{2\pi}{\sum_i 1/\sigma_i^2}}\right),
\end{equation}
where $\alpha_i=\mu_{obs}(z_i)-25-5\log_{10}[H_0 d_L(z_i)]$, and $\mathbf{p}$ denotes
the fitting parameters in the model. When using the SN Ia data,
the radiation term can be neglected because its contribution is negligible.

In addition to the Constitution SN Ia data, we  use the BAO
distance measurements from the oscillations in the distribution of
galaxies. From the BAO observation of the galaxy power spectra,
Percival {\it et al} measured the distance ratio
\begin{equation}
\label{dz} d_{z}= \frac{r_{s}(z_{d})}{D_{V}(z)}
\end{equation}
at two redshifts $z=0.2$ and $z=0.35$ to be
$d_{0.2}^{obs}=0.1905\pm 0.0061$, and $d_{0.35}^{obs}=0.1097\pm
0.0036$, respectively \cite{wjp}. Here the effective distance is
\begin{equation}
\label{dvdef}
D_V(z)=\left[\frac{d_L^2(z)}{(1+z)^2}\frac{z}{H(z)}\right]^{1/3},
\end{equation}
the drag redshift $z_d$ is fitted as \cite{dw}
\begin{equation}
\label{zdfiteq} z_d=\frac{1291(\Omega_m
h^2)^{0.251}}{1+0.659(\Omega_m h^2)^{0.828}}[1+b_1(\Omega_b
h^2)^{b_2}],
\end{equation}
\begin{eqnarray}
\label{b1eq} b_1=0.313(\Omega_m h^2)^{-0.419}[1+0.607(\Omega_m
h^2)^{0.674}],  \quad b_2=0.238(\Omega_m h^2)^{0.223},
\end{eqnarray}
the comoving sound horizon is
\begin{equation}
\label{rshordef} r_s(z)=\int_z^\infty \frac{c_s(z)dz}{E(z)},
\end{equation}
the sound speed $c_s(z)=1/\sqrt{3[1+\bar{R_b}/(1+z)}]$, and
$\bar{R_b}=3\Omega_b h^2/(4\times2.469\times10^{-5})$. To use
these BAO data, we calculate
\begin{equation}
\label{baochi2} \chi^2_{BAO2}(\mathbf{p},\Omega_b h^2, h)=\Delta
x_i {\rm Cov_{1}}^{-1}(x_i,x_j)\Delta x_j,
\end{equation}
where $x_i=(d_{z=0.2},d_{z=0.35})$, $\Delta x_i=x_i-x_i^{obs}$ and
Cov$_{1}(x_i,x_j)$ is the covariance matrix for the two parameters $d_{0.2}$ and $d_{0.35}$
\cite{wjp}. Besides the model parameters $\mathbf{p}$, we need
to add two more parameters $\Omega_b h^2$ and $\Omega_m h^2$ when we use the BAO data.
In \cite{wjp}, they used the
priors of $\Omega_b h^2=0.02273\pm 0.00061$ and $\Omega_c h^2=0.1099\pm 0.0063$.

From the measurement of the radial (line-of-sight) BAO scale in the
galaxy power spectra, the cosmological parameters were determined
from the measured values of
\begin{equation}
\label{deltaz} \Delta z_{BAO}(z)=\frac{H(z)r_{s}(z_{d})}{c}
\end{equation}
at two redshifts $z=0.24$ and $z=0.43$, which are
$\Delta z_{BAO}(z=0.24)=0.0407\pm 0.0011$ and $\Delta z_{BAO}(z=0.43)=0.0442\pm0.0015$, respectively \cite{eg}.
Therefore, we add $\chi^{2}$ with
\begin{equation}
\label{baochi3} \chi^2_{BAOz}(\mathbf{p},\Omega_b h^2,
h)=\left(\frac{\Delta z_{BAO}(0.24)-0.0407}{0.0011}\right)^2
+\left(\frac{\Delta z_{BAO}(0.43)-0.0442}{0.0015}\right)^2.
\end{equation}
When we add these BAO data to the fitting, we also need to use the
parameters $\Omega_b h^2$ and $\Omega_m h^2$. The values $\Omega_b
h^2=0.02273\pm 0.0066$ and $\Omega_m h^2=0.1329\pm 0.0064$ were
used in \cite{eg}.

In addition to the above two BAO data sets, the BAO $A$ parameter \cite{dje} is usually used.
The BAO $A$ parameter is defined as
\begin{equation}
\label{para1}
A=\sqrt{\Omega_{m}}\frac{H_0 D_V(z=0.35)}{z=0.35}=\frac{\sqrt{\Omega_{m}}}{0.35}\left[\frac{0.35}{E(0.35)}\frac{1}{|\Omega_{k}|}{\rm
sinn}^2\left(\sqrt{|\Omega_{k}|}\int_0^{0.35}
\frac{dz}{E(z)}\right)\right]^{1/3},
\end{equation}
and it was measured to be $A=0.493\pm 0.017$ \cite{bar},
so we add $\chi^{2}$ with
\begin{equation}
\label{baochi1}
\chi^2_{BAOa}(\mathbf{p})=\left(\frac{A-0.439}{0.017}\right)^2.
\end{equation}
Note that the BAO $A$ parameter depends on the model parameters
$\mathbf{p}$ only; it does not depend on the baryon density
$\Omega_b h^2$ and the Hubble constant $h$. Although the radiation
density depends on $h$, the contribution to the Hubble parameter
$E(z)$ is negligible at the redshift $z=0.35$, so we can neglect
the radiation component when we use the BAO $A$ data.

Both the SN Ia and the BAO data measure the distance up to redshit
$z<2$; we need to consider the distance at hight redshift in order
to determine the property of dark energy. Therefore, we implement
the WMAP5 data. To use the full WMAP5 data, we need to add some
more parameters which depend on inflationary models, and this will
limit our ability to constrain dark energy models. So we only use
the WMAP5 measurements of the derived quantities, such as the
shift parameter $R(z^{*})$, the acoustic scale $l_A(z^{*})$ and
the decoupling redshift $z^{*}$, to obtain
\begin{equation}
\label{cmbchi} \chi^2_{CMB}=\Delta x_i {\rm
Cov_{2}}^{-1}(x_i,x_j)\Delta x_j,
\end{equation}
where the three parameters $x_i=(R(z^{*}),\ l_A(z^{*}),\ z^{*})$, $\Delta
x_i=x_i-x_i^{obs}$ and Cov$_{2}(x_i,x_j)$ is the covariance matrix
for the three parameters \cite{kdn}. The shift parameter $R$ is
expressed as
\begin{equation}
\label{shift}
R(z^{*})=\frac{\sqrt{\Omega_{m}}}{\sqrt{|\Omega_{k}|}}{\rm
sinn}\left(\sqrt{|\Omega_{k}|}\int_0^{z^{*}}\frac{dz}{E(z)}\right)=1.710\pm
0.019.
\end{equation}
 The acoustic scale $l_A$ is
\begin{equation}
\label{ladefeq} l_A(z^{*})=\frac{\pi
d_L(z^{*})}{(1+z^{*})r_s(z^{*})}=302.1\pm0.86,
\end{equation}
and the decoupling redshift $z^{*}$ is fitted by \cite{hs}
\begin{eqnarray}
\label{zstareq} z^{*}=1048[1+0.00124(\Omega_b
h^2)^{-0.738}][1+g_1(\Omega_m h^2)^{g_2}] =1090.04\pm 0.93,
\end{eqnarray}
\begin{equation}
g_1=\frac{0.0783(\Omega_b h^2)^{-0.238}}{1+39.5(\Omega_b
h^2)^{0.763}},\quad g_2=\frac{0.560}{1+21.1(\Omega_b h^2)^{1.81}}.
\end{equation}
In \cite{kdn}, it was found that $\Omega_b h^2=0.02273\pm 0.00062$
and $h=0.719^{+0.026}_{-0.027}$.

The SN Ia data, the BAO data and the WMAP5 data use the distance
measurement to determine the cosmological parameters. To get the
distance scale, we need to integrate the equation of state
parameter $w(z)$ twice, so the process of double integration
smoothes out the variation of equation of state parameter $w(z)$
of dark energy. To alleviate the problem, we add the Hubble
parameter $H(z)$ data. The Hubble parameter $H(z)$ at nine
different redshifts was obtained from the differential ages of
passively evolving galaxies in \cite{hz1}, and three more Hubble
parameter data $H(z=0.24)=76.69\pm2.32$, $H(z=0.34)=83.8\pm2.96$
and $H(z=0.43)=86.45\pm3.27$ were determined recently in
\cite{hz2}. Therefore, we add these $H(z)$ data to $\chi^2$:
\begin{equation}
\label{hzchi}
\chi^2_H(\mathbf{p}, h)=\sum_{i=1}^{12}\frac{[H(z_i)-H_{obs}(z_i)]^2}{\sigma_{hi}^2},
\end{equation}
where $\sigma_{hi}$ is the $1\sigma$ uncertainty in the $H(z)$
data.
The model parameters $\mathbf{p}$ are determined by applying the
maximum likelihood method of $\chi^{2}$ fit. We use the
publicly available MINUIT code for minimization and contour
calculation \cite{fm}. Basically,
The model parameters are determined by minimizing
\begin{equation}
\label{chi2min}
\chi^2=\chi^2_{sn}+\chi^2_{Baoa}+\chi^2_{Bao2}+\chi^2_{Baoz}+\chi^2_{CMB}+\chi^2_H.
\end{equation}
For the convenience of numerical fitting, we take $\Omega_b
h^2=0.02273$ determined from the WMAP5 data \cite{kdn}.
For the Hubble constant $h$, two different values were observed. The Hubble key project found that $h=0.72\pm 0.08$,
and recently Riess {\it et al} obtained $h=0.742\pm 0.036$ by using a differential distance ladder method \cite{riess09}.
To account for the uncertainty of the Hubble constant, we treat it as
a free parameter and then fix it at its best fit value.

\section{$\Lambda$CDM model with curvature}

For the cosmological constant, the equation of state parameter
$w=p/\rho=-1$, and the energy density $\rho_\Lambda$ is a
constant. In a curved $\Lambda$CDM model, the curvature term
$k\neq 0$, ordinary pressureless dust matter, radiation  and the
cosmological constant contribute to the total energy. The
Friedmann equation is
\begin{eqnarray}
\label{elcdm}
E(z)=\frac{H(z)}{H_0}=[\Omega_{k}(1+z)^2+\Omega_{m}(1+z)^3
 +\Omega_{r}(1+z)^4+\Omega_{\Lambda}]^{1/2},
\end{eqnarray}
where the Hubble constant $H_{0}=100h  km s^{-1} Mpc^{-1}$,
$\Omega_m=(8\pi G\rho_m)/(3H^2_0)$ is the current matter
component, the current radiation component $\Omega_{r}=(8\pi
G\rho_r)/(3H^2_0)=4.1736\times 10^{-5}h^{-2}$ \cite{kdn}, the
current curvature component $\Omega_k=-k/(a_0^2 H_0^2)$  and
$\Omega_{\Lambda}=1-\Omega_{m}-\Omega_{k}-\Omega_{r}$. In this
model, we have two parameters $\mathbf{p}=(\Omega_m,\ \Omega_k)$
and one nuisance parameter $h$. For the fitting to the SN Ia data,
the contribution to the Hubble expansion from the radiation is
negligible and we usually neglect the radiation term.

\begin{figure}[htp]
$\begin{array}{cc}
\includegraphics[width=0.46\textwidth]{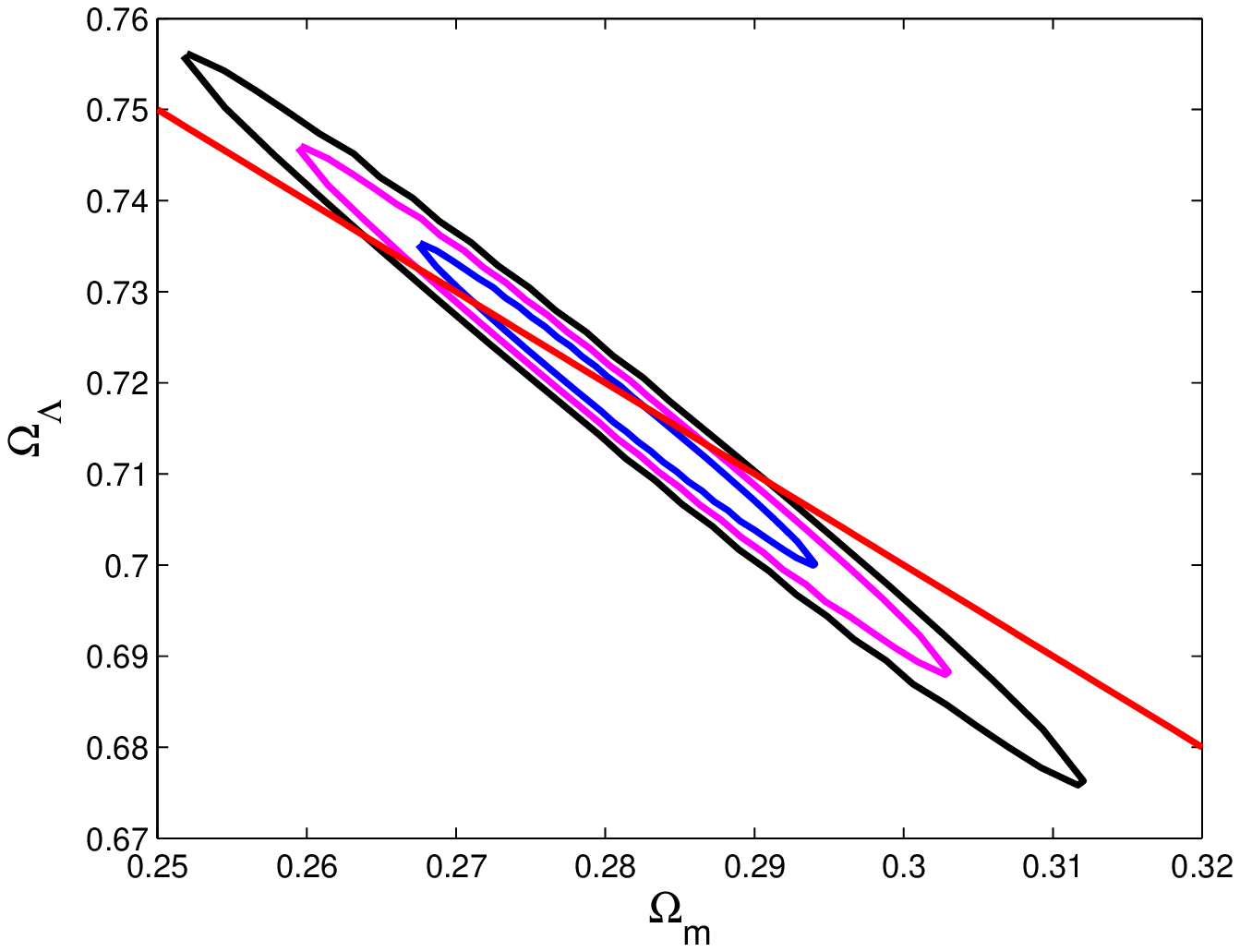}&
\includegraphics[width=0.46\textwidth]{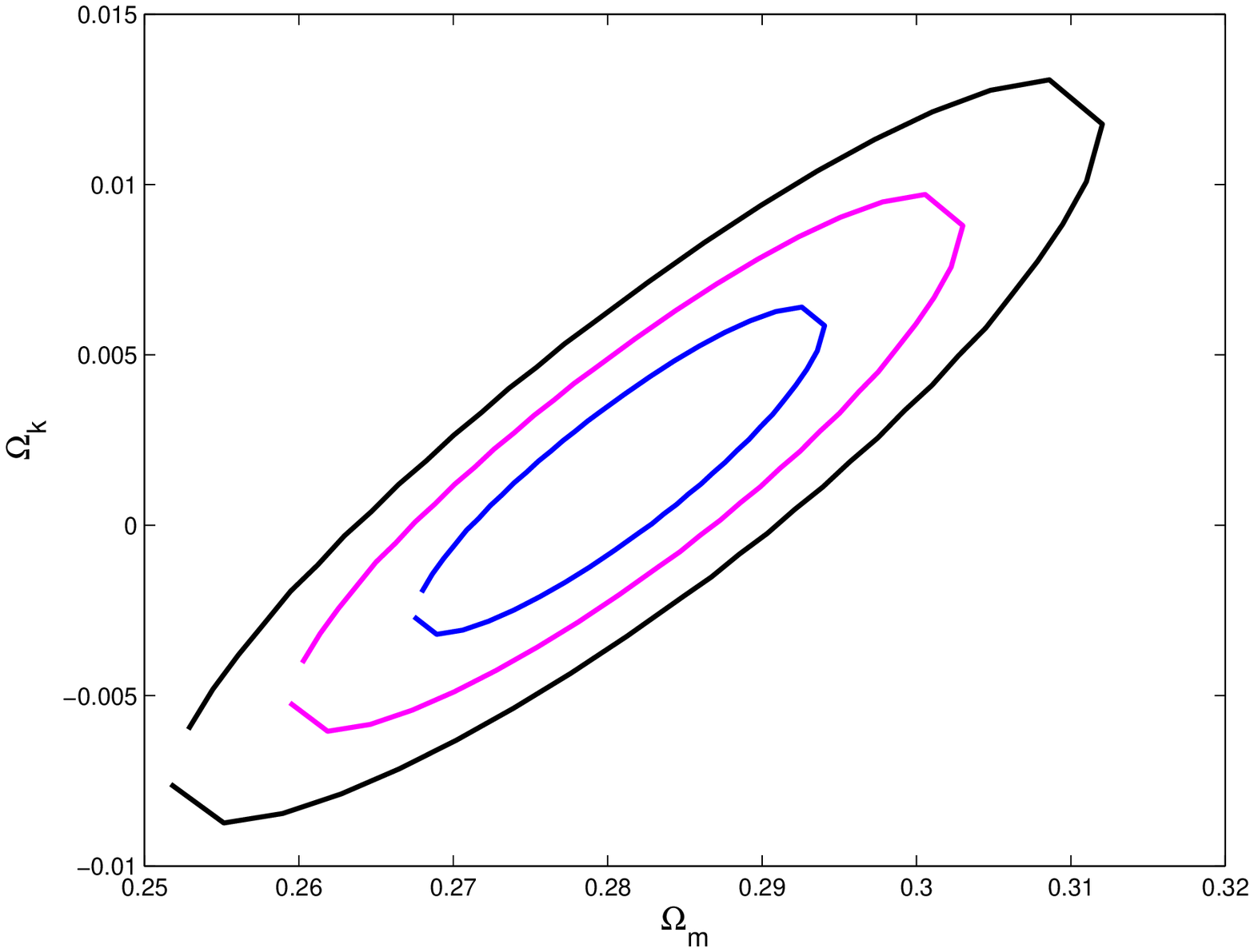}
\end{array}$
\caption{The $1\sigma$, $2\sigma$ and $3\sigma$ joint contour plots of
$\Omega_{m}$ and $\Omega_\Lambda$ ($\Omega_{k}$) for the curved $\Lambda$CDM model. The straight line
in the left panel denotes the flat $\Lambda$CDM model.}
\label{LCDM1}
\end{figure}

By fitting the $\Lambda$CDM model to the above observational data,
we get $\chi^2=482.13$, $\Omega_m=0.280^{+0.014}_{-0.013}$  and
$\Omega_k=0.001\pm 0.005$. By fixing the nuisance parameter $h$ at
its best fit value $h=0.702$, we obtain the contours of $\Omega_m$
and $\Omega_k$. The joint contour plots of $\Omega_{m}$ and
$\Omega_{k}$ or $\Omega_\Lambda$ are shown in figure \ref{LCDM1}.
Compared with WMAP5 fitting results \cite{kdn}, we find that the
current data make a little improvement on the constraints of
$\Omega_{m}$ and $\Omega_{k}$. The improvement is due to more SN
Ia and BAO data in addition to the $H(z)$ data. The result tells
us that the flat $\Lambda$CDM model is consistent with current
observational data at the $1\sigma$ level.

\section{CPL parametrization with curvature}

\begin{figure}[htp]
$\begin{array}{cc}
\subfigure[]{\includegraphics[width=0.4\textwidth]{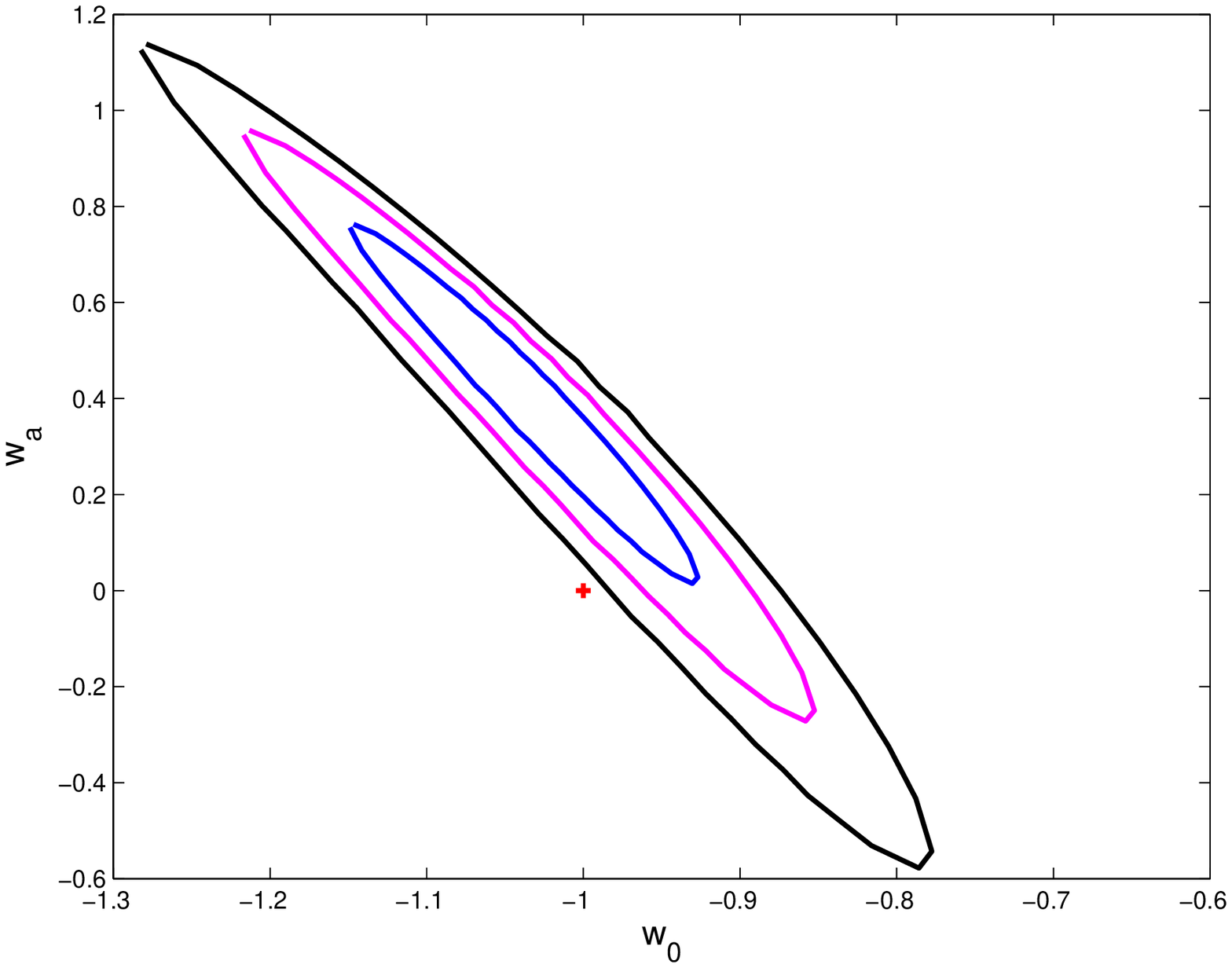}
\label{cpl1:a}}&
\subfigure[]{ \includegraphics[width=0.4\textwidth]{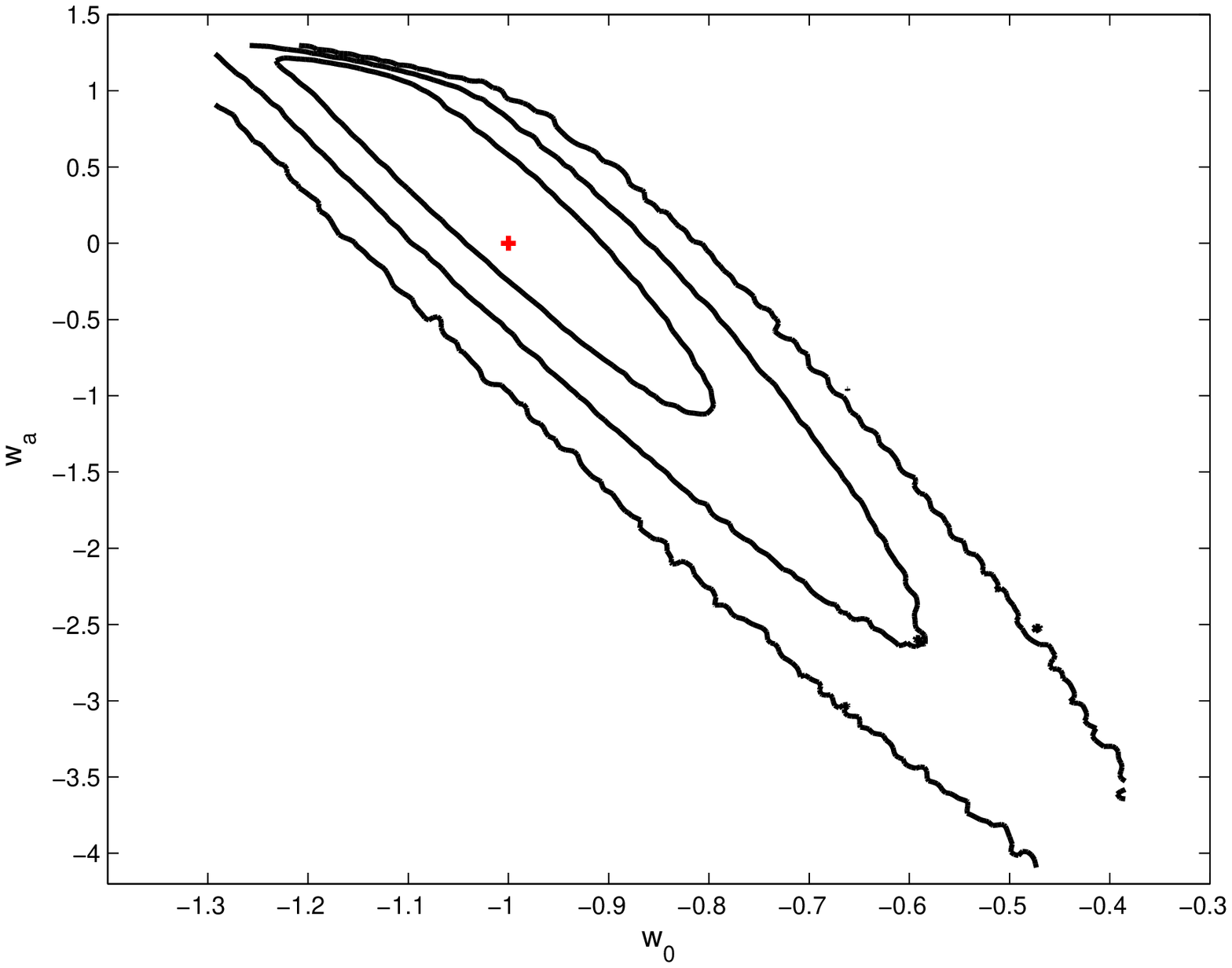}
\label{cpl1:b}}
\end{array}$
\caption{The $1\sigma$, $2\sigma$ and $3\sigma$ contour plots of
$w_0$ and $w_a$ for the curved CPL parametrization. '+' denotes
the point corresponding to the $\Lambda$CDM model. (a) Joint
contours of $w_0$ and $w_a$ by fixing the other parameters at
their best fit values. (b) Marginalized contours of $w_0$ and
$w_a$ obtained from the MCMC method.} \label{CPL1}
\end{figure}

In order to investigate the equation of state of dark energy
for a curved universe by observational data, in this section we study the
popular CPL parametrization \cite{cpl}
\begin{equation}
\label{lind} w(z)=w_0+\frac{w_a z}{1+z}.
\end{equation}
The dimensionless Hubble parameter including the contributions from dark energy,
ordinary pressureless dust matter and radiation is
\begin{eqnarray}
\label{cplez}
E(z)=\frac{H(z)}{H_0}=(\Omega_{k}(1+z)^2+\Omega_{m}(1+z)^3
 +\Omega_{r}(1+z)^4+\Omega_{DE})^{1/2},
\end{eqnarray}
where the dimensionless dark energy density is
\begin{eqnarray}
\label{deneq}
\Omega_{DE}(z)=(1-\Omega_{m}-\Omega_{k}-\Omega_{r})
\times(1+z)^{3(1+w_0+w_a)}\exp[-3w_az/(1+z)].
\end{eqnarray}
In this model, we have four model parameters
$\mathbf{p}=(\Omega_{m},\ \Omega_{k},  \ w_0, \ w_a)$. By applying
the observational data discussed in the previous section to the
CPL model, we are able to get the observational constraint on the
model parameters $\mathbf{p}=(\Omega_{m},\ \Omega_{k},  \ w_0, \
w_a)$. The best fit is $\chi^2=481.64$, $\Omega_m=0.278$,
$\Omega_k=0.006$, $w_0=-1.04$, $w_a=0.42$ and $h=0.70$. By fixing
the parameters $\Omega_m$, $\Omega_k$ and $h$ at their best fit
values, we obtain the contours of $w_0$ and $w_a$ and they are
shown in figure \ref{cpl1:a}. From figure \ref{cpl1:a}, we see
that the $\Lambda$CDM model is excluded by the observational data
at more than $3\sigma$ level. As we discussed in the previous
section, we see that the $\Lambda$CDM model is consistent with the
observational data.  The totally different conclusions suggest
that the simple $\chi^2$ error estimation by fixing other
parameters at their best fit values has some drawbacks because we
neglect the correlation effects of the other parameters. The
degeneracy between parameters was not considered in the above
method. When the parameters are strongly correlated, the error of
some parameters will be under-estimated if we fix the other
parameters at their best fit values. To verify this point, we
apply the MCMC method to constrain the parameter space $\bm{p}$
and the nuisance parameters $h$ and $\Omega_b h^2$.  Our MCMC code
\cite{gong08} is based on the publicly available package COSMOMC
\cite{cosmomc}. By using the MCMC method, we get $\chi^2=481.27$;
the marginalized $1\sigma$ errors are
$\Omega_m=0.279^{+0.015}_{-0.008}$,
$\Omega_k=0.005^{+0.006}_{-0.011}$, $w_0=-1.05^{+0.23}_{-0.06}$
and $w_a=0.5^{+0.3}_{-1.5}$. These results are summarized in table
\ref{table1}. The marginalized $1\sigma$, $2\sigma$ and $3\sigma$
contour plots of $w_0$ and $w_a$ are shown in figure \ref{cpl1:b}.
From figure \ref{cpl1:a} and \ref{cpl1:b}, we find that the
marginalized $1\sigma$ contour obtained by using the MCMC method
includes the $3\sigma$ contour in figure \ref{cpl1:a}. From figure
\ref{cpl1:b}, we see that the $\Lambda$CDM model is consistent
with the CPL model at $1\sigma$ level.

\begin{figure}[htp]
$\begin{array}{cc}
\subfigure[]{\includegraphics[width=0.4\textwidth]{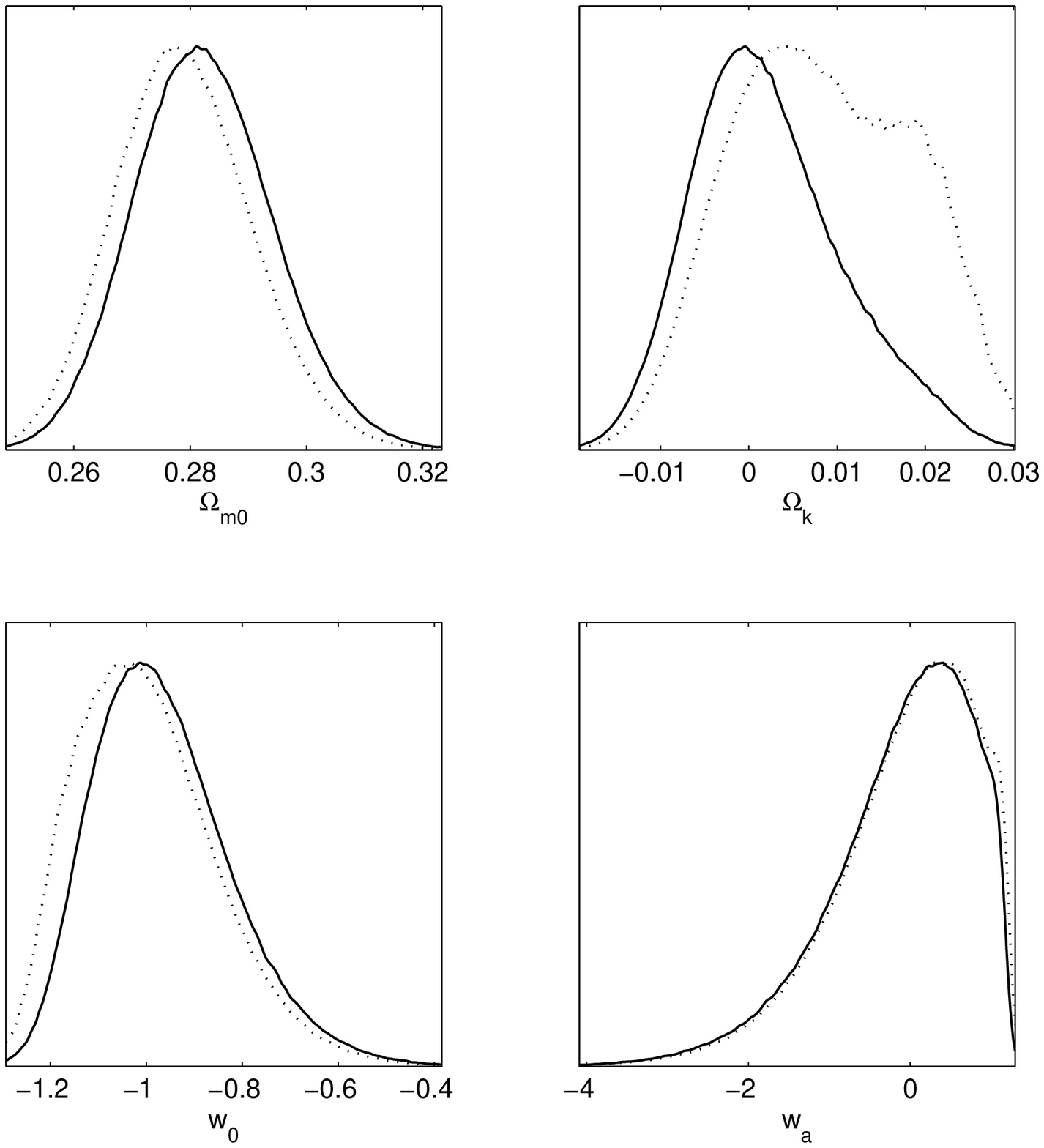}
\label{cpl2:a}}&
\subfigure[]{ \includegraphics[width=0.4\textwidth]{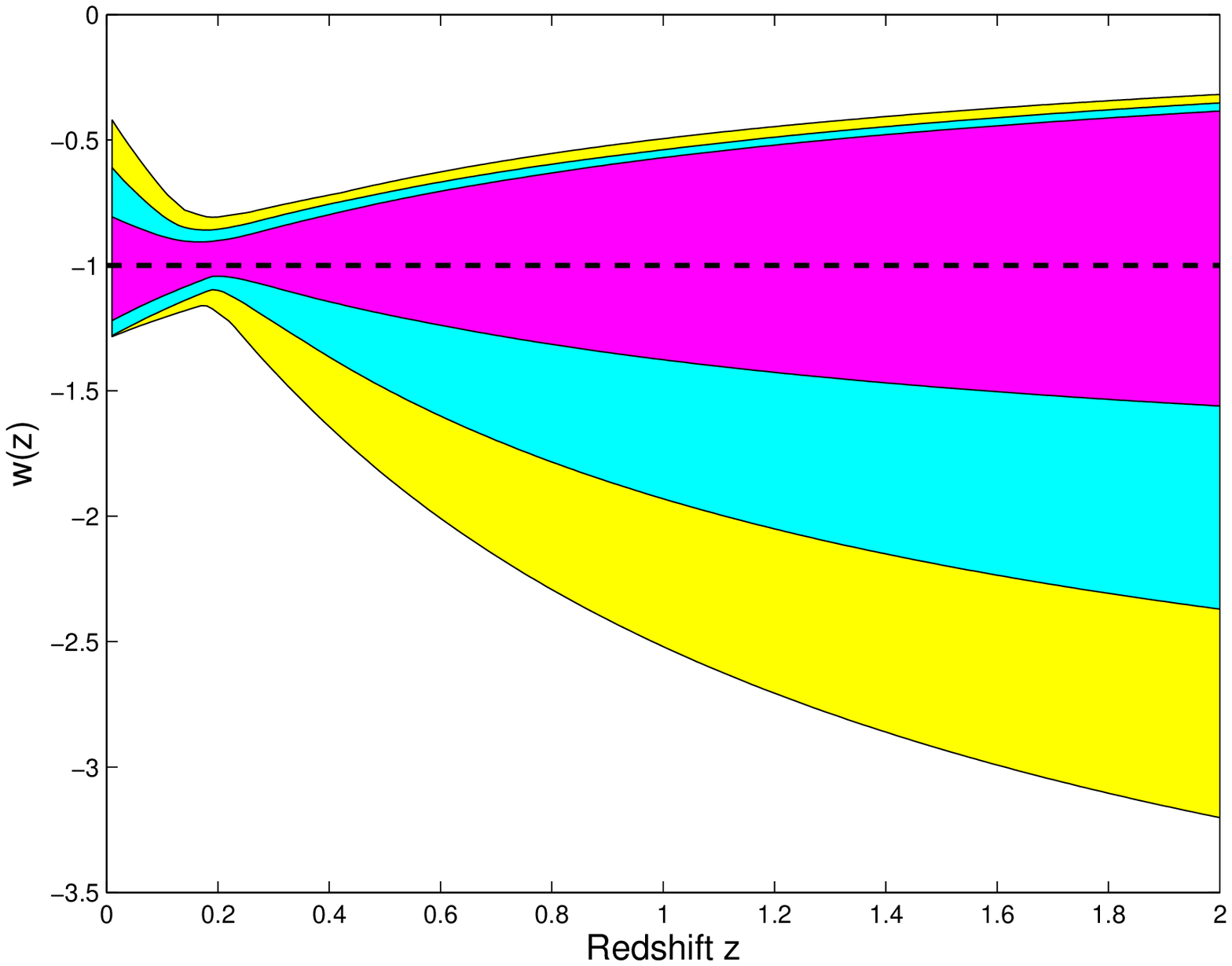}
\label{cpl2:b}}
\end{array}$
\caption{The marginalized distributions of the CPL model
parameters $\mathbf{p}$ are shown in (a). The solid lines are
marginalized probabilities and the dotted lines are mean
likelihoods. (b) Reconstructed evolution of $w(z)$ and the shaded
areas are $1\sigma$, $2\sigma$ and $3\sigma$ errors.} \label{CPL2}
\end{figure}

\begin{table*}
\begin{center}
\caption{The marginalized estimates of the model parameters in CPL and JBP models.} \label{table1}
\begin{tabular}{|l|c|c |c|c|c |c|}
\hline Model  & $\chi^{2}$ & $\Omega_{m}$ &  $\Omega_{k}$ &
$w_{0}$ & $w_{a}$\\ \hline
 CPL
 & 481.27
 & $0.279^{+0.015}_{-0.008}$
  & $0.005^{+0.006}_{-0.011}$
  & $-1.05^{+0.23}_{-0.06}$
  & $0.47^{+0.28}_{-1.45}$\\ \hline
  JBP
 & 481.46
 & $0.281^{+0.015}_{-0.01}$
  & $0.000^{+0.007}_{-0.006}$
  & $-0.96^{+0.25}_{-0.18}$
  & $-0.6^{+1.9}_{-1.6}$\\ \hline
\end{tabular}
\end{center}
\end{table*}

The marginalized distributions of the model parameters are shown
in figure \ref{cpl2:a}. The solid lines are marginalized
probabilities and the dotted lines represent mean likelihoods.
From figure \ref{cpl2:a}, we find that the likelihood of
$\Omega_k$ has a local maximum around $\Omega_k\sim 0.02$. Even we
take $\Omega_k=0.02$, the value of $\chi^2$ is not far from the
minimum value of $\chi^2$.  Due to the degeneracy between model
parameters, if we fix the other model parameters at their best fit
values, then the joined contours of $w_0$ and $w_a$ are
under-estimated, and the conclusion drawn from the under-estimated
contours is not reliable. The results in figure \ref{cpl1:b} and
\ref{cpl2:a} verify this point. By using the marginalized contours
of $w_0$ and $w_a$, we reconstruct the evolution of $w(z)$ in
figure \ref{cpl2:b}. From figure \ref{cpl2:b}, we find that $w(z)<
0$ at more than $3\sigma$ confidence level up to redshift $z=2$,
and the $\Lambda$CDM model is consistent with the CPL model at the
$1\sigma$ level. To account for the correlations between model
parameters, we need to use the marginalized probability. To see
whether this happens only for the CPL model, we analyze the JBP
model in the next section.

\section{JBP parametrization with curvature}

\begin{figure}[htp]
$\begin{array}{cc}
\subfigure[]{\includegraphics[width=0.4\textwidth]{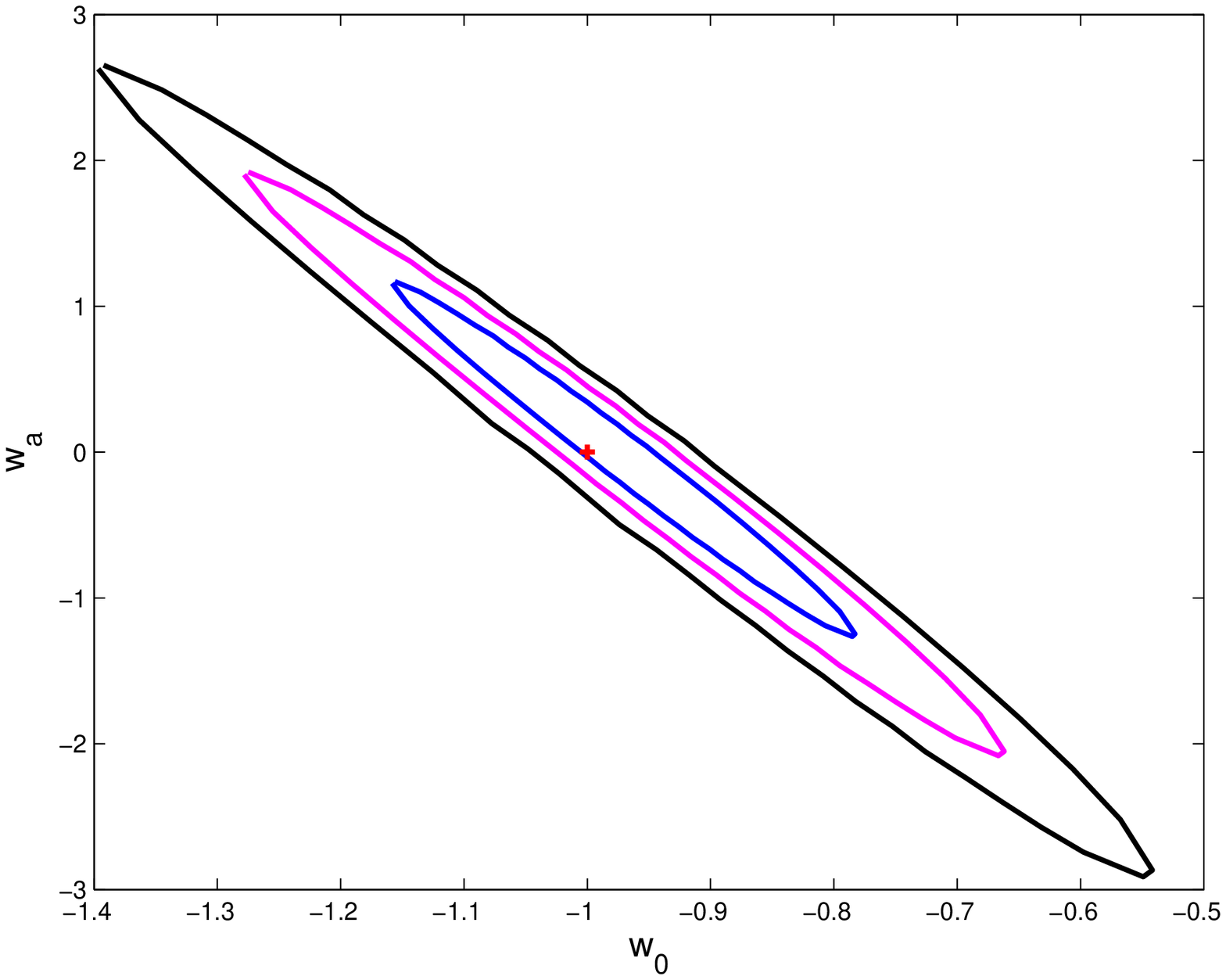}
\label{jbp2:a}}&
\subfigure[]{ \includegraphics[width=0.4\textwidth]{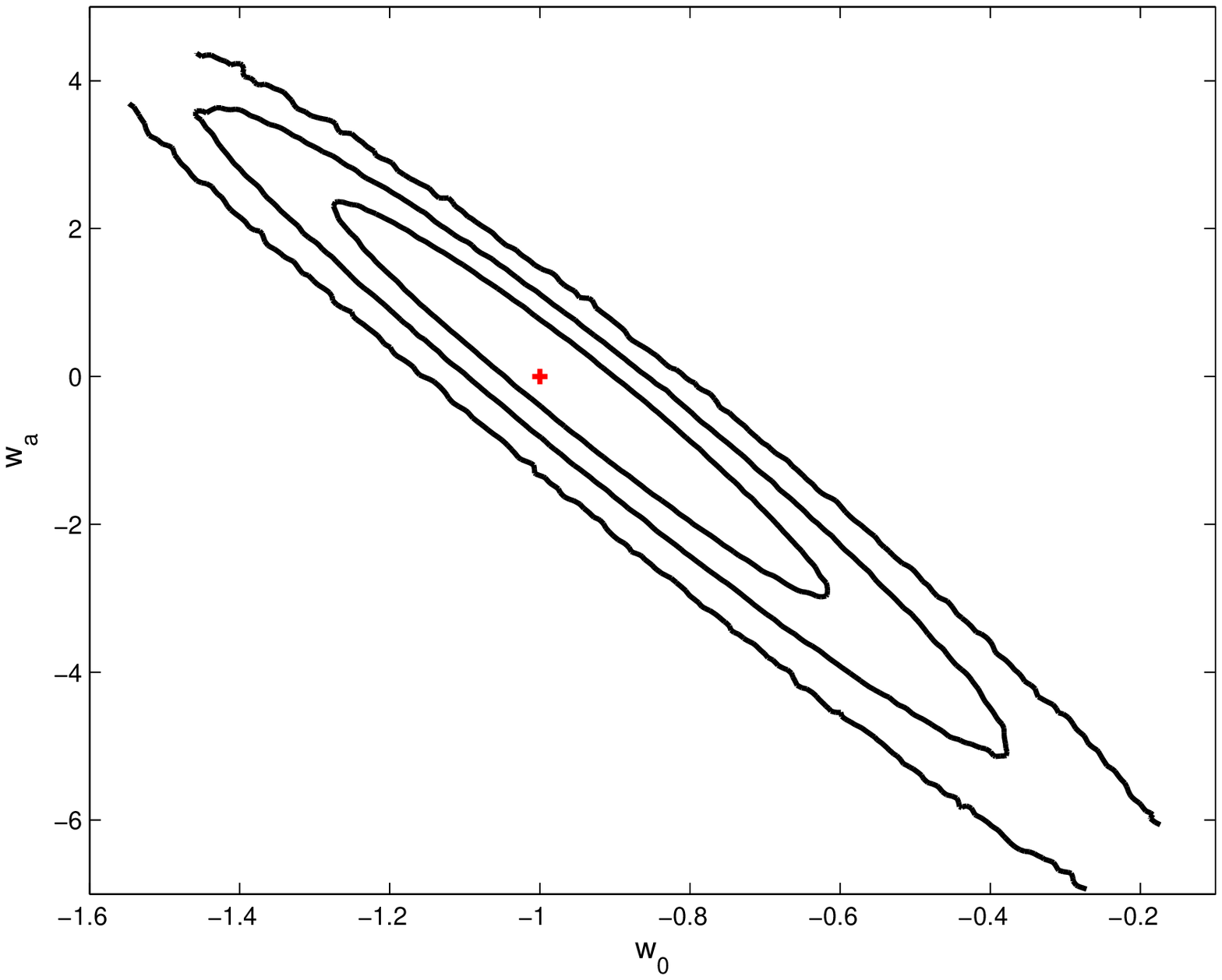}
\label{jbp2:b}}
\end{array}$
\caption{The $1\sigma$, $2\sigma$ and $3\sigma$ contour plots of
$w_0$ and $w_a$ for the curved JBP parametrization. '+' denotes
the point corresponding to the $\Lambda$CDM model. (a) Joint
contours of $w_0$ and $w_a$ by fixing the other parameters at
their best fit values. (b) Marginalized contours of $w_0$ and
$w_a$ obtained from the MCMC method.} \label{JBP2}
\end{figure}

In this section, we consider the JBP parametrization \cite{jbp}
for dark energy with the equation
 of state in the form below
\begin{equation}
\label{jbpw} w(z)=w_0+\frac{w_a z}{(1+z)^2}\ .
\end{equation}
The corresponding dimensionless dark energy density is then
\begin{eqnarray}
\label{deneq1}
\Omega_{DE}(z)=(1-\Omega_{m}-\Omega_{k}-\Omega_{r})
\times(1+z)^{3(1+w_0)}\exp\left[3w_az^2/2(1+z)^2\right].
\end{eqnarray}
In this model, we also have four parameters
$\mathbf{p}=(\Omega_{m},\ \Omega_{k},  \ w_0, \ w_a)$. We first
use the simple $\chi^2$ method to fit the model. The best fit is
$\chi^2=481.84$, $\Omega_m=0.281$, $\Omega_k=0.0015$, $w_0=-0.97$,
$w_a=-0.03$  and $h=0.70$. By fixing the parameters $\Omega_m$,
$\Omega_k$ and $h$ at their best fit values, we obtain the
contours of $w_0$ and $w_a$ and they are shown by the solid lines
in figure \ref{jbp2:a}. Unlike the CPL model, the $\Lambda$CDM
model is consistent with the JBP model at the $1\sigma$ level. To
verify this conclusion, we also apply the MCMC method to the JBP
model. By using the MCMC method, we get $\chi^2=481.46$; the
marginalized $1\sigma$ errors are
$\Omega_m=0.281^{+0.015}_{-0.01}$,
$\Omega_k=0.000^{+0.007}_{-0.006}$, $w_0=-0.96^{+0.25}_{-0.18}$
and $w_a=-0.6^{+1.9}_{-1.6}$. These results are summarized in
table \ref{table1}. The marginalized $1\sigma$, $2\sigma$ and
$3\sigma$ contour plots of $w_0$ and $w_a$ are shown in figure
\ref{jbp2:b}. From figure \ref{jbp2:a} and \ref{jbp2:b}, we see
that the contours of $w_0$ and $w_a$ are consistent although the
constraints from the MCMC method are a little larger because we
consider the correlations among all the parameters in the MCMC
method. The $\Lambda$CDM model is also consistent with the JBP
model at the $1\sigma$ level.

\begin{figure}[htp]
$\begin{array}{cc}
\subfigure[]{\includegraphics[width=0.4\textwidth]{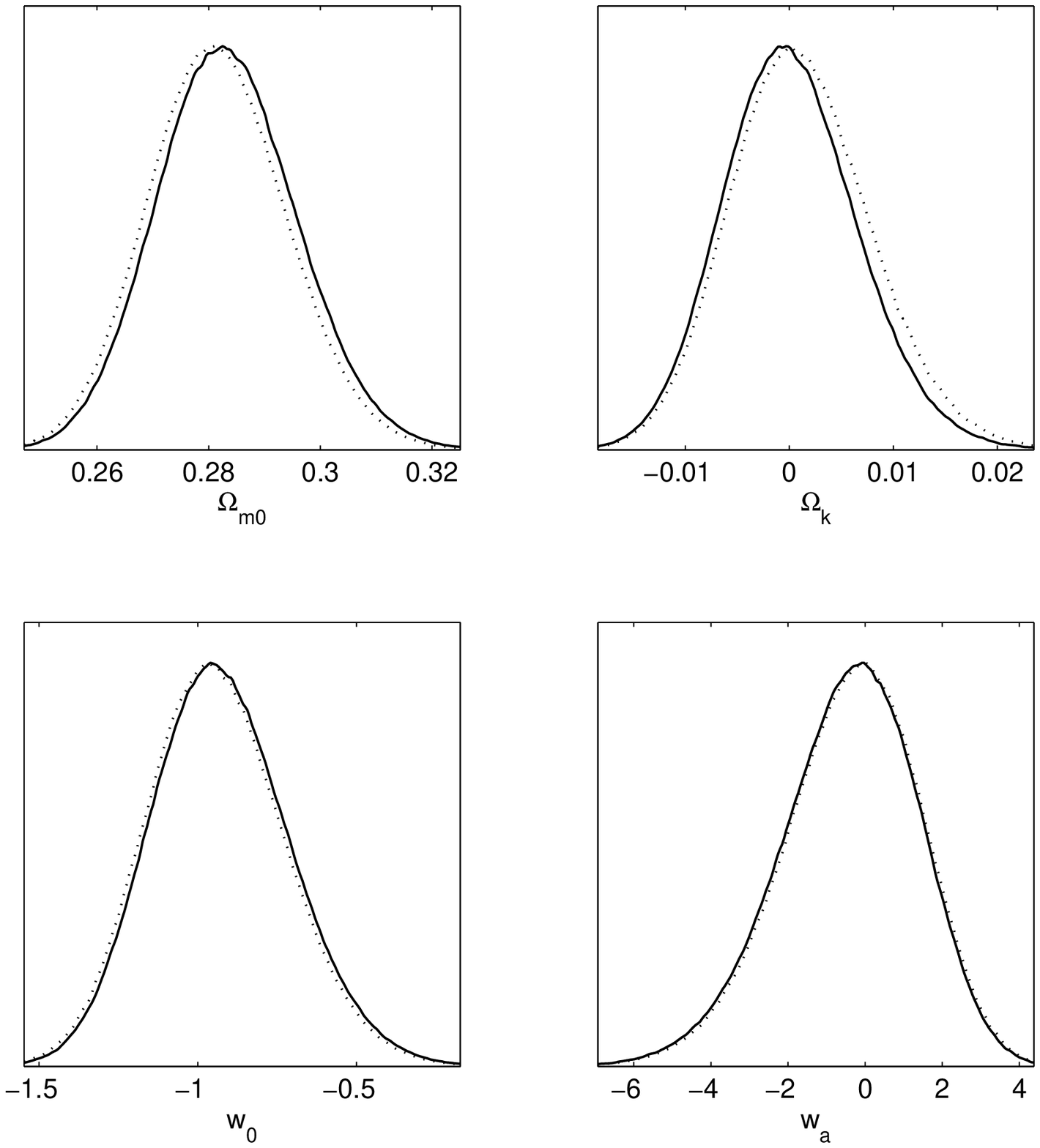}
\label{jbp1:a}}&
\subfigure[]{ \includegraphics[width=0.4\textwidth]{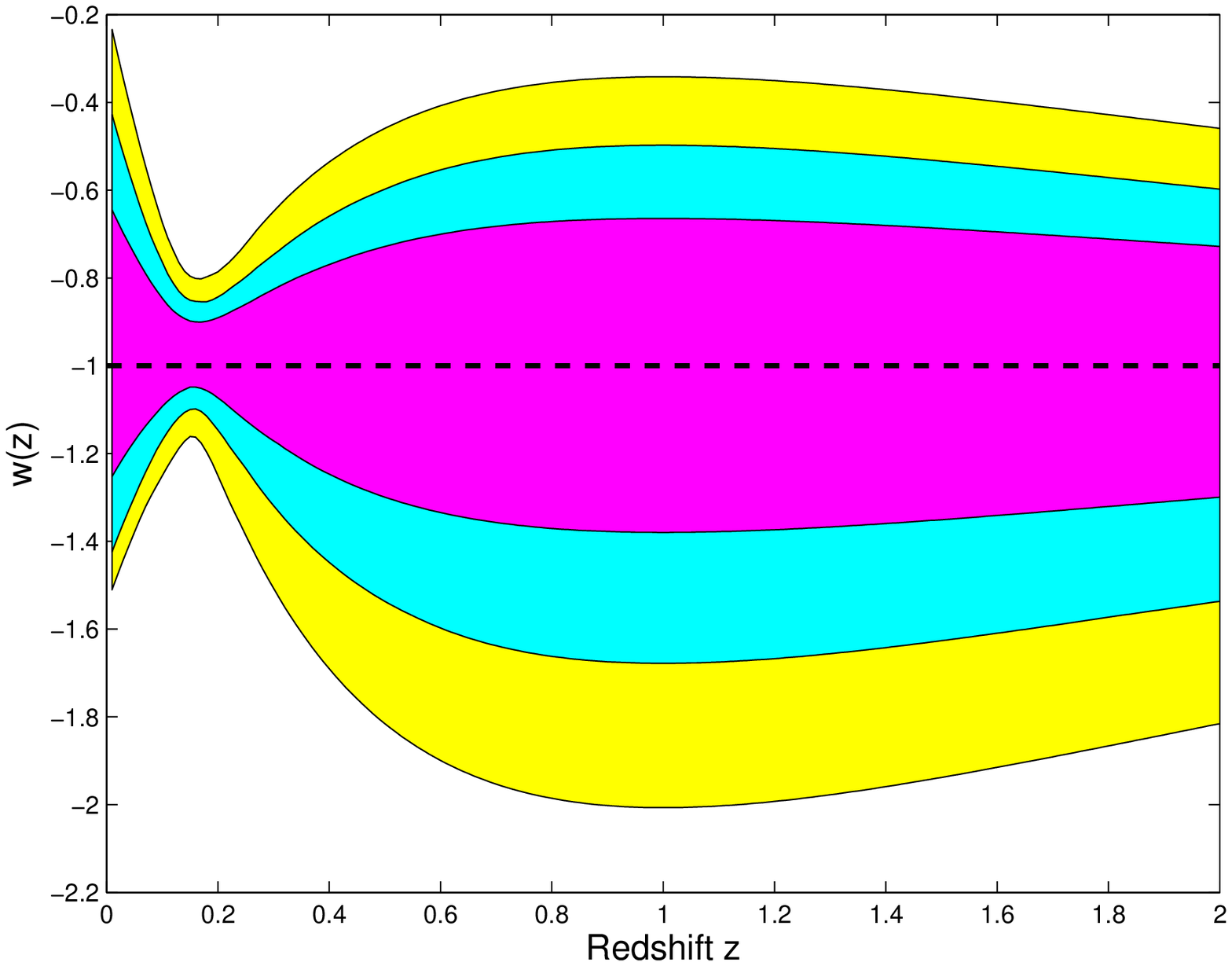}
\label{jbp1:b}}
\end{array}$
\caption{The marginalized distributions of the JBP model
parameters $\mathbf{p}$ are shown in (a). The solid lines are
marginalized probabilities and the dotted lines are mean
likelihoods. (b) Reconstructed evolution of $w(z)$ and the shaded
areas are $1\sigma$, $2\sigma$ and $3\sigma$ errors.} \label{JBP1}
\end{figure}

The marginalized distributions of the model parameters
$\mathbf{p}$ are shown in figure \ref{jbp1:a}. The solid lines are
marginalized probabilities and the dotted lines represent mean
likelihoods. From figure \ref{jbp1:a}, we find that the
probability distributions of the parameters are more or less
Gaussian. By using the marginalized contours of $w_0$ and $w_a$,
we reconstruct the evolution of $w(z)$ in figure \ref{jbp1:b}.
From figure \ref{jbp1:b}, we find that $w(z)< -0.2$ at more than
$3\sigma$ confidence level up to redshift $z=2$, and the
$\Lambda$CDM model is consistent with the JBP model at the
$1\sigma$ level.

\section{Conclusions}

Applying the simple $\chi^2$ method, we fitted the CPL and JBP
models to the combined SN Ia, BAO, WMAP5 and $H(z)$ data, and
obtained the constraint on the property of dark energy. In both
CPL and JBP models, there are four parameters
$\mathbf{p}=(\Omega_{m},\ \Omega_{k},  \ w_0, \ w_a)$. When we
apply the BAO and WMAP5 data, we need to add two more parameters
$\Omega_b h^2$ and $h$. If we make the joint error analysis, we
have six parameters and it will be hard to get a good joint
constraint on all these parameters. Therefore, we take $\Omega_b
h^2=0.02273$, and find out the best fit values of the parameters
$\mathbf{p}$ and $h$ which minimize $\chi^2$; then we fix the
parameters $\Omega_m$, $\Omega_k$ and $h$ at their best fit values
to obtain the joint constraints on $w_0$ and $w_a$. For the CPL
model, the contours of $w_0$ and $w_a$ (see figure \ref{cpl1:a})
show that the $\Lambda$CDM model is excluded at more than
$3\sigma$ level. The JBP model is consistent with the $\Lambda$CDM
model at the $1\sigma$ level. Since we get the contours of $w_0$
and $w_a$ by fixing the other parameters at their best fit values,
we neglect the correlation effects of the parameters and the
conclusion based on this method may not be reliable. To confirm
this, we use the MCMC method to analyze the CPL and JBP models and
obtain the marginalized probabilities of the parameters. For the
CPL model, the probability distributions of $\Omega_k$ and $w_a$
are skew distributions, and the marginalized $1\sigma$ errors are
$\Omega_m=0.279^{+0.015}_{-0.008}$,
$\Omega_k=0.005^{+0.006}_{-0.011}$, $w_0=-1.05^{+0.23}_{-0.06}$
and $w_a=0.5^{+0.3}_{-1.5}$. In the CPL model, the probability
distributions of $\Omega_k$ has a local maximum in addition to a
global maximum. The uncertainties in $\Omega_k$ and the
degeneracies between $\Omega_k$, $w_0$ and $w_a$ lead to
under-estimation of the error contours of $w_0$ and $w_a$ if we
fix $\Omega_k$ at its global best fit value, and the wrong
conclusion that the $\Lambda$CDM model is excluded at more than
$3\sigma$ level. However, this does not happen for the JBP model.
For the JBP model, the parameters have Gaussian distributions, and
the marginalized $1\sigma$ errors are
$\Omega_m=0.281^{+0.015}_{-0.01}$,
$\Omega_k=0.000^{+0.007}_{-0.006}$, $w_0=-0.96^{+0.25}_{-0.18}$
and $w_a=-0.6^{+1.9}_{-1.6}$.

In summary, in addition to use the usual SN Ia, BAO $A$ or BAO
distance ratio, and WMAP data, we also use the radial BAO
measurements and the $H(z)$ data to fit the CPL and JBP models. We
find that the equation of state parameter of dark energy $w(z)<0$
at more than $3\sigma$ level in the redshift range $0\le z\le 2$, 
and the flat $\Lambda$CDM model is consistent with
the current observational data at the $1\sigma$ level.
Furthermore, we find that we need to do the marginalized analysis
to estimate the errors of the model parameters.

\ack

NP was partially supported by the project A2008-58 of the
Scientific Research Foundation of Chongqing University of Posts
and Telecommunications, and the NNSF of China under grant no
10947178. YG was partially supported by the NNSF key project of
China under grant no 10935013, the National Basic Research Program
of China under grant no 2010CB833004, and the Natural Science
Foundation Project of CQ CSTC under grant no 2009BA4050. ZZ was
partially supported by the NNSF Distinguished Young Scholar
project under Grant no 10825313, and the National Basic Research
Program of China under grant no 2007CB815401.


\section*{References}
\begin{thebibliography}{acc1}
\bibitem{acc1} Riess A G {\it et al} 1998 {\it Astron. J.} {\bf 116} 1009
\bibitem{acc2} Perlmutter S {\it et al} 1999 {\it Astrophy. J.} {\bf 517} 565
\bibitem{quint} Wetterich C 1988 {\it Nucl. Phys.} B {\bf 302} 668 \\
Ratra B and  Peebles P J E 1988 {\it \PR} D {\bf 37} 3406 \\
Caldwell R R, Dave R and Steinhardt P J 1998 {\it \PRL} {\bf 80} 1582
\bibitem{phantom} Caldwell R R 2002 {\it \PL} B {\bf 545} 23
\bibitem{k} Armendariz-Picon C, Damour T and Mukhanov V  1999 {\it \PL} B {\bf
458} 209
\bibitem{tachyonic} Padmanabhan T 2002 {\it \PR} D {\bf 66} 021301
\\
Bagla J S, Jassal H K, and Padmanabhan T 2003 {\it \PR} D {\bf67} 063504
\bibitem{Feng:2004ad} Feng B, Wang X L and Zhang X M 2005 {\it \PL}  B {\bf 607} 35
\bibitem{hessence} Wei H, Cai R G, and Zeng D F 2005 {\it \CQG} {\bf 22} 3189\\
Wei H, Cai R G, and Zeng D F 2005 {\it \PR} D {\bf 72} 123507
\bibitem{chaplygin} Kamenshchik A Y, Moschella U and Pasquier V 2001 {\it \PL} B
{\bf 511} 265 \\
Bento M C, Bertolami O and Sen A A 2002 {\it \PR} D {\bf 66} 043507
\bibitem{holo} Hsu S D H 2004 {\it \PL} B {\bf 594} 13 \\
Li M 2004 {\it \PL} B {\bf 603} 1
\bibitem{fr} Capozziello  S 2002, {\it Int. J. Mod. Phys.} D {\bf11} 483
\\
Nojiri S and Odintsov S D 2003 {\it \PR} D {\bf 68} 123512\\
Hu W and Sawicki I 2007 {\it \PR} D {\bf76} 064004
\bibitem{dgp} Dvali  G, Gabadadze G and Porrati M 2000 {\it \PL} B {\bf 485} 208
\bibitem{cpl} Chevallier M and Polarski D 2001  {\it Int. J. Mod.
Phys.} D {\bf 10} 213 \\
Linder E V 2003 {\it \PRL} {\bf 90} 091301
\bibitem{star} Shafieloo A, Sahni V and Starobinsky A A 2009 {\it \PR} D {\bf 80} 101301
\bibitem{huang} Huang Q G, Li M, Li X D and Wang S 2009 {\it \PR} D {\bf 80} 083515
\bibitem{cai} Cai R G, Su Q P and Zhang H-B 2010 {\it J. Cosmol. Astropart. Phys.}
JCAP04(2010)012
\bibitem{corray9} Serra  P {\it et al} 2009 {\it \PR} D {\bf 80} 121302
\bibitem{gong} Gong Y G, Cai R G, Chen Y and Zhu Z-H 2010 {\it J. Cosmol. Astropart. Phys.} JCAP01(2010)019
\bibitem{gong10} Gong Y G, Wang B and Cai R G 2010 {\it J. Cosmol. Astropart. Phys.} JCAP04(2010)019
\bibitem{consta}  Hicken M {\it et al} 2009 {\it Astrophys. J.} {\bf 700} 1097
\bibitem{bar}  Reid B A {\it et al} 2010 {\it Mon. Not. R. Astron. Soc.} {\bf 404} 60
\bibitem{wjp}  Percival W J {\it et al} 2010  {\it Mon. Not. R. Astron. Soc.} {\bf 401} 2148
\bibitem{eg} Gazta\~{n}aga E, Miquel R and S\'{a}nchez E 2009 {\it \PRL} {\bf 103} 091302
\bibitem{kdn} Komatsu E {\it et al} 2009 {\it Astrophys. J. Suppl. Ser.} {\bf 180} 330
\bibitem{hz1} Simon J, Verde L and Jimenez R 2005 {\it \PR} D {\bf 71} 123001
\bibitem{hz2} Gazta\~{n}aga E, Cabr\'{e} A and Hui L 2009  {\it Mon. Not. R. Astron. Soc.} {\bf 399} 1663
\bibitem{jbp}  Jassal H K, Bagla J S and Padmanabhan T 2005
{\it Mon. Not. Roy. Astron. Soc.} {\bf 356} L11
\bibitem{gong08} Gong Y G, Wu Q and Wang A 2008 {\it Astrophys. J.} {\bf 681} 27
\bibitem{dw} Eisenstein D J and Hu W 1998 {\it Astrophys. J.} {\bf 496} 605
\bibitem{dje} Eisenstein  D J {\it et al} 2005 {\it Astrophys. J.} {\bf 633} 560
\bibitem{hs} Hu W and Sugiyama N 1996 {\it Astrophys. J.} {\bf 471} 542
\bibitem{fm} James F and Roos M 1975 {\it Comput. Phys. Commun.} {\bf 10} 343
\bibitem{hst} Freedman W L {\it et al} 2001 {\it Astrophys. J.} {\bf 553} 47
\bibitem{riess09} Riess A G {\it et al} 2009 {\it Astrophys. J.} {\bf 699} 539
\bibitem{cosmomc} Lewis A and Bridle S 2002 {\it \PR} D {\bf 66} 103511


\end{thebibliography}
\end{document}